\documentclass[3p,times,twocolumn]{elsarticle}

\usepackage{ecrc}
\usepackage{amssymb}

\usepackage{graphics}
\usepackage{graphicx}
\usepackage{epsfig}

\volume{00}

\firstpage{1}

\journalname{Physica B}

\runauth{V.F. Nasretdinova et al.}

\jid{physb}

\jnltitlelogo{Physica B}

\usepackage{amssymb}
\begin{document}

\begin{frontmatter}


 \title{Indium doping-induced change in the photoconduction spectra of {\it o-}TaS$_3$}
\author[label1]{V.F. Nasretdinova\corref{cor1},}
 \author[label2]{E.B. Yakimov,}
 \author[label1,label3]{S.V. Zaitsev-Zotov}
 \address[label1]{Kotel'nikov Institute of Radioengineering and Electronics of RAS, Mokhovaya 11, bld.7, 125009 Moscow, Russia}
 \address[label2]{Institute of Microelectronics Technology and
High Purity Materials of RAS, Chernogolovka, Moscow distr., 142432 Russia}
\address[label3]{Moscow Institute of Physics and Technology, 141700 Dolgoprudny, Russia} 
 \cortext[cor1]{Tel: +7 (495) 629 33 94; FAX:+7 (495) 629 36 78; e-mail: venera@cplire.ru (V. Nasretdinova)}

%\title{}


%\author{}

%\address{}

\begin{abstract}
Impurities and defects are known to affect the properties of the charge density wave (CDW) state but the influence of impurities on the density of states inside the Peierls gap remains largely unexplored. Here we present an experimental study of the effect of indium impurities on photoconduction spectra of CDW compound orthorhombic TaS$_3$. We use the temperature diffusion method to introduce indium into a sample from preliminary attached In contacts. The concentration of In after 23 hours of diffusion is found to be nonuniform and strongly dependent on the distance to the contacts. The diffusion affects the spectral range 0.15-0.25 eV, increasing the photoconduction amplitude linearly with diffusion time. The optical gap value obtained from the measurements is $2\Delta = 0.25$~eV and the tail of states below $2\Delta$ is associated with the impurities in agreement with the T\"{u}tt\"{o}-Zawadowski theory. Diffusion-induced modification of current-voltage characteristics and decrease of the Peierls temperature are also observed. Neither changes in photoconduction spectra nor in the Peierls transition temperature of the control sample with Au contacts are found. 
\end{abstract}

\end{frontmatter}

\section{Introduction}
Impurities and defects strongly affect properties of the charge density wave (CDW) state in quasi-one dimensional (q-1D) conductors. The Peierls transition temperature, $T_P$ and broad range of phenomena coupled to the pinning and sliding of the CDW depend on impurities \cite{monceau}.

However consistent understanding of the effect of impurities on the CDW gap is still lacking. This effect appears to be quite distinct from that in usual semiconductors. For example, STM study of NbSe$_3$ demonstrates that doping can either decrease (Fe) or increase (Co) the gap value $2\Delta$ to the tens of percent already at dilute concentrations \cite{Coleman}.
Discrete states inside the CDW gap were predicted for a single impurity site in 1D metallic chain and tail of states was predicted for a substantial concentration of weak-pinning impurities \cite{tutto}. Intra-gap states in {\it o-}TaS$_3$ were observed in a recent photoconduction spectral study \cite{TaS3} but were not identified with impurities. A tail in bolometric response was observed \cite{brill} but the authors argued whether it originates from impurities or from other effects.    

Investigation of gap features in relatively small samples of TaS$_3$ that are non-suitable for STM or ARPES techniques is possible however in photoconduction spectra study. A photoconduction spectrum close to fundamental absorption edge is defined mainly by electronic absorption spectrum of the sample which allows one to exclude phonon absorption crucial in bolometric response studies, as well as absorption by thermally excited electrons and holes.  In this paper we report the results of the studies of the indium doping effect on the CDW gap in q-1D conductor {\it o-}TaS$_3$ investigated by means of the photoconduction spectroscopy.

\section{Experimental}
We employed the method developed by Gill \cite{Gill} for NbSe$_3$ for introduction indium into {\it o-}TaS$_3$ samples. Namely, Gill found that  heating of NbSe$_3$ with indium contacts for one hour at $120^o$C increases the bulk concentrations of In by approximately 0.1 atomic \%  throughout the 2-mm length sample owing to high coefficient of In diffusion. Indium was presumed to intercalate the crystals. 

\begin{figure}
\includegraphics[width=7cm]{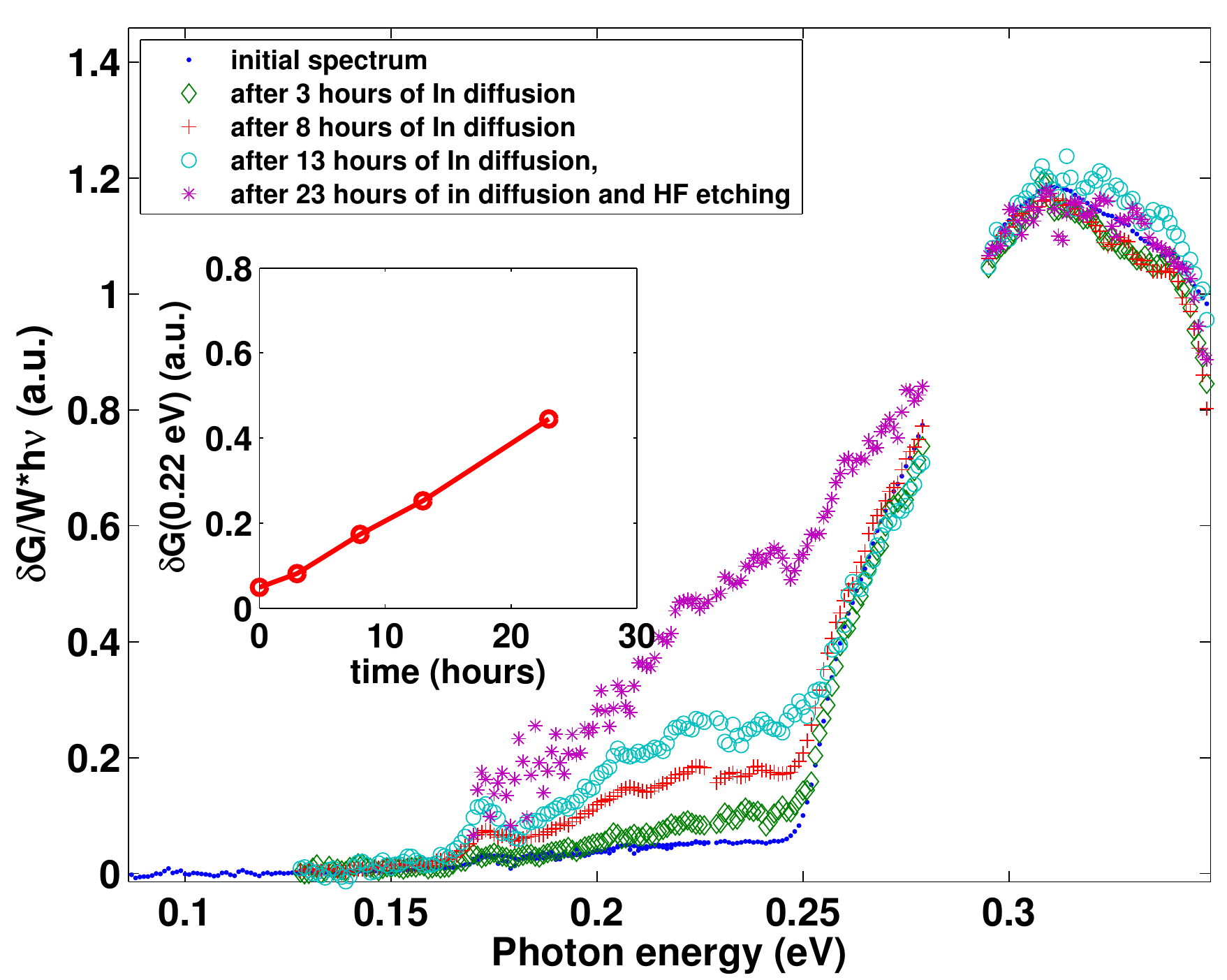}
\caption{Evolution of the photoconduction spectrum of the specimen of the {\it o-}TaS$_3$ with consequent In processng. Five spectra correspond to different time of In processng; y-axes shows photoconductivity normalized by power of incident light, beside this all spectra are normalized to unity at maximum for convenience. T = 40 K, voltage applied to the sample is below $E_{th} \times l$ where $l$ is the sample's length and $E_{th}$ --- treshold field for the onset of the CDW motion. Inset: the relative value of photoconductivity at 0.22 eV versus the total processing time.}
\label{Fig.1}
\end{figure}

We studied two samples of {\it o-}TaS$_3$ denoted below as samples \#G and \#Y one of which (sample \#G) was the control sample. The sample \#G  was chosen from the batch synthesized by R.E.~Thorne and the sample \#Y was chosen from the batch of crystals grown in IRE RAS by direct reaction of high-purity Ta and S in a quartz tube at a temperature gradient 680-710$^o$C for 10 days. The sample \#Y contacts were made of indium by the cold soldering. The width of each In contact was $d \sim 200{\rm \ }\mu$m, the distance between contacts --- $387{\rm \ } \mu$m, the sample cross-section --- $2{\rm \ } \mu$m$^2$. During the heating processing the sample \#Y was kept at 120$^o$C under the constant Ar flow for 1 up to 10 hours. Full data set which consists of photoconduction spectra, $IV$ curve sets and temperature dependencies of resistance was measured between the processing procedures. In addition, we also performed heating in the air and did not found any difference in the character of spectral changes. Therefore the respective spectra are also included in the data set (see Fig.~\ref{Fig.1}).
The same effect of the heating in Ar atmosphere and on the air indicates absence of the noticeable contribution in photoconduction from the surface oxidation.

When the total duration of the processing reached 23 hours we performed etching of the sample \#Y in concentrated hydrofluoric acid (HF) \cite{HF} to get rid of possible surface states caused by possible surface oxidation. The procedure was to dip the sample in HF for 60 seconds then clean it in a flow of a distilled water. The full data set was collected before and after etching.

The control sample \#G had the gold contacts deposited by the thermal evaporation and no indium was used for attachment of the sample to the measurement circuit. We applied the same heating processing and collected the full data set for the sample \#G before and after the processing.

The temperature-dependent variation of conduction and $IV$ curves were measured by the DC technique in two-contact configuration in the voltage-controlled regime. We used the same photoconduction measurements setup as in our earlier works \cite{TaS3}. It consists of light-induced conduction variation measurement by the standard lock-in technique in the two-contact configuration with the sample placed in cryostat in He heat-exchange gas at the normal pressure. The temperature was controlled with the accuracy of 0.1~K. All the photoconduction spectra presented in this work represent photoconductivity normalized by the number of incident photons, $\delta G \hbar\omega/W$, where $\delta G = \delta I/V$ is photoconduction, $\delta I$ is measured photo-current, $V$, $W$, $\hbar \omega$ are the voltage, light intensity and photon energy respectively.

\section{Results}
\subsection{Photoconduction spectra}
Fig.~\ref{Fig.1} shows the set of photoconduction (PC) spectra obtained on different stages of the processing. The initial spectrum (empty circles) shows a tail of states starting at $\hbar\omega^* = 0.15$~eV followed by a sharp growth of photoconduction at $\hbar\omega = 0.25$~eV. The latter we ascribe to the Peierls gap edge for the reasons explained below.
The gradual increase of the photoconductivity at photon energies $0.15$~eV $< \hbar\omega <0.25$~eV with growth of the total processing time with respect to photoconductivity at $\hbar\omega = 0.25$~eV are clearly seen on Fig.~\ref{Fig.1}. The inset in Fig.~\ref{Fig.1} shows the dependence of the photoconduction at $\hbar\omega = 0.22$~eV  on the total processing time.  Processing shorter than 13 hours affects only the amplitude of the photoconduction tail at low photon energies $\hbar\omega < 0.25$~eV but 23 hours processing smears out the gap edge by $\delta \Delta \sim 50$~meV. Such a behavior is consistent with previous results for {\it o-}TaS$_3$ spectra: relatively thick fresh-cleaved samples demonstrate pronounced gap edge, i.e. photoconductivity drop at values close to $\hbar\omega = 0.25$~eV, whereas thin and/or old samples always show very smeared gap edge.

Changes in PC spectra proportional to the time of In introduction at $\hbar\omega < 0.25$~eV let us conclude that  0.25 eV corresponds to the band-band transition over the CDW gap $2\Delta$. But onset of the photoconduction is close to $\hbar\omega^*=0.15$~eV matching the value obtained from the transport measurements for this particular sample $2\Delta_{tr} \approx 0.15$ eV where $\Delta_{tr} = 900$~ K is resistivity activation energy, as well as reported transport values \cite{cdwreview,TaS3hall}. Such a discrepancy between the onset of PC and the sharp drop of PC, i.e. tail of states, is characteristic even for nominally pure {\it o-}TaS$_3$ PC spectra \cite{TaS3} and may include contributions from different effects. One possibility is the imperfect nesting \cite{Yama,Maki} resulting in order parameter modulation in k-space and indirect gap $2 \Delta_i$ with phonon-assisted transitions at $2 \Delta_i <\hbar\omega <2\Delta$~eV. The other contribution may arise from self-trapped states typical for q-1D systems \cite{braz}. In addition, T\"{u}tt\"{o} and Zawadowski obtained \cite{tutto} that impurities in q-1D conductor in the  weak-pinning limit would cause the region inside the gap to be filled by states if the distance between impurities is smaller than CDW amplitude coherence length $\xi_0$. This would shift the gap edge to the new value $2\Delta' = 2\Delta (1-t^2)^{1/2}$ where $t=T/v_F$ is impurity backscattering amplitude over the Fermi velocity. Considering $\xi_{0} \sim 10$~nm \cite{tutto} we obtain the corresponding concentration of impurities $n_{imp}$ of the order $10^{18}$~cm$^{-3} \sim 10^{-4}$~at.\% which is consistent with the values for nominally pure samples. Taking shifted gap edge value $2\Delta'= 0.15$~eV and direct gap value $2\Delta = 0.25$~eV one gets $t=0.6$ which is close to $t=0.7$ obtained in Ref.~\cite{brill} and consistent with the case of weak pinning. We conclude that the observed photoconductivity increase with the In introduction corresponds to the latter case. 
However, the density of states obtained by considering both imperfect nesting and the modulation of the gap value in k-space gives remarkable fit to the spectral features at $\hbar\omega > 2\Delta$, see \cite{ecrys2014}.

\begin{figure}
\includegraphics[width=7cm]{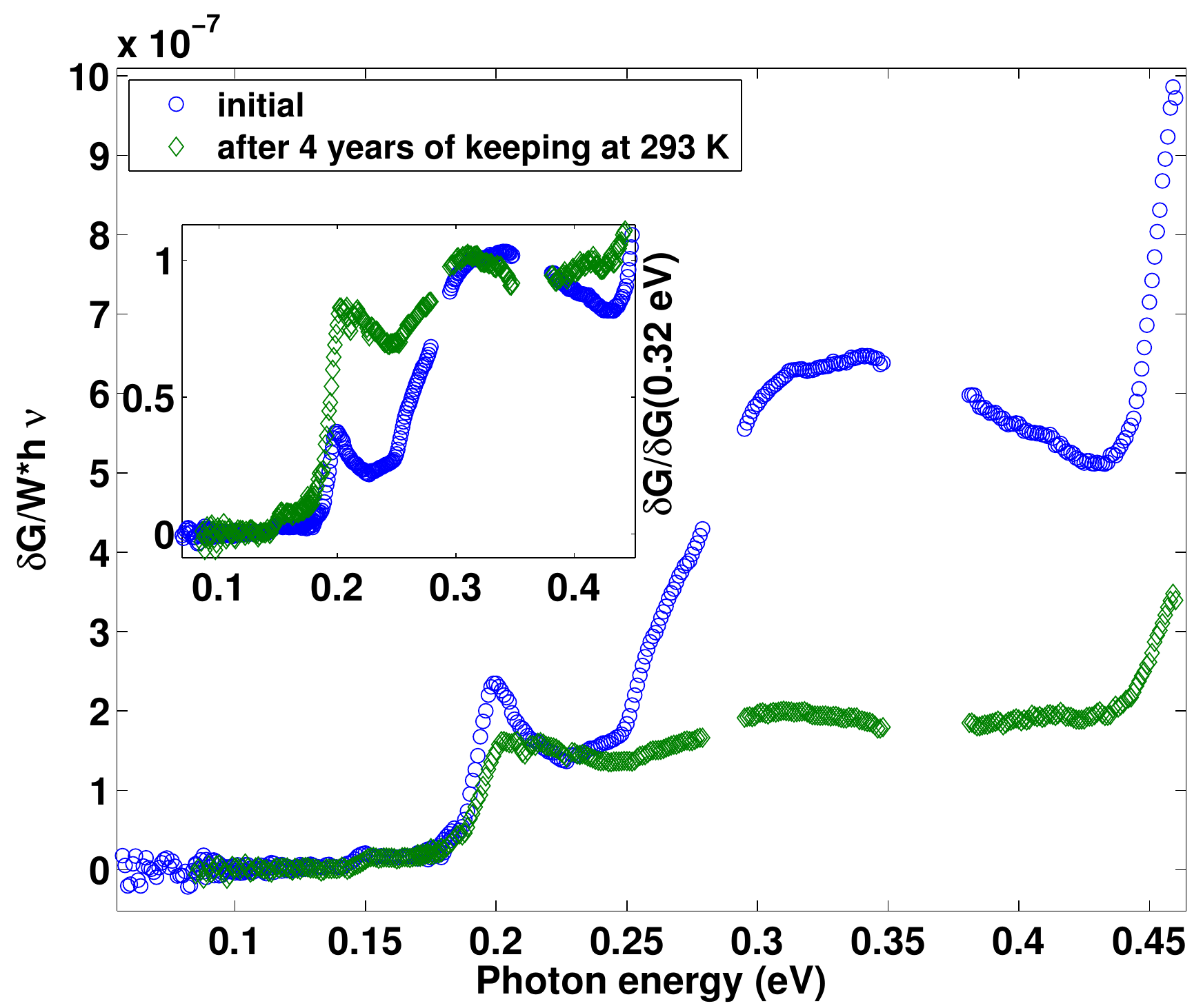}
\caption{Initial photoconduction spectrum of the {\it o-}TaS$_3$ sample \#Z with In contacts and spectrum of the same sample measured after few years of keeping at room temperature. Inset shows same spectra normalized at maximum by unity. $T = 40$~K}
\label{Fig.2}
\end{figure}

\begin{figure}[h]
\includegraphics[width=7cm]{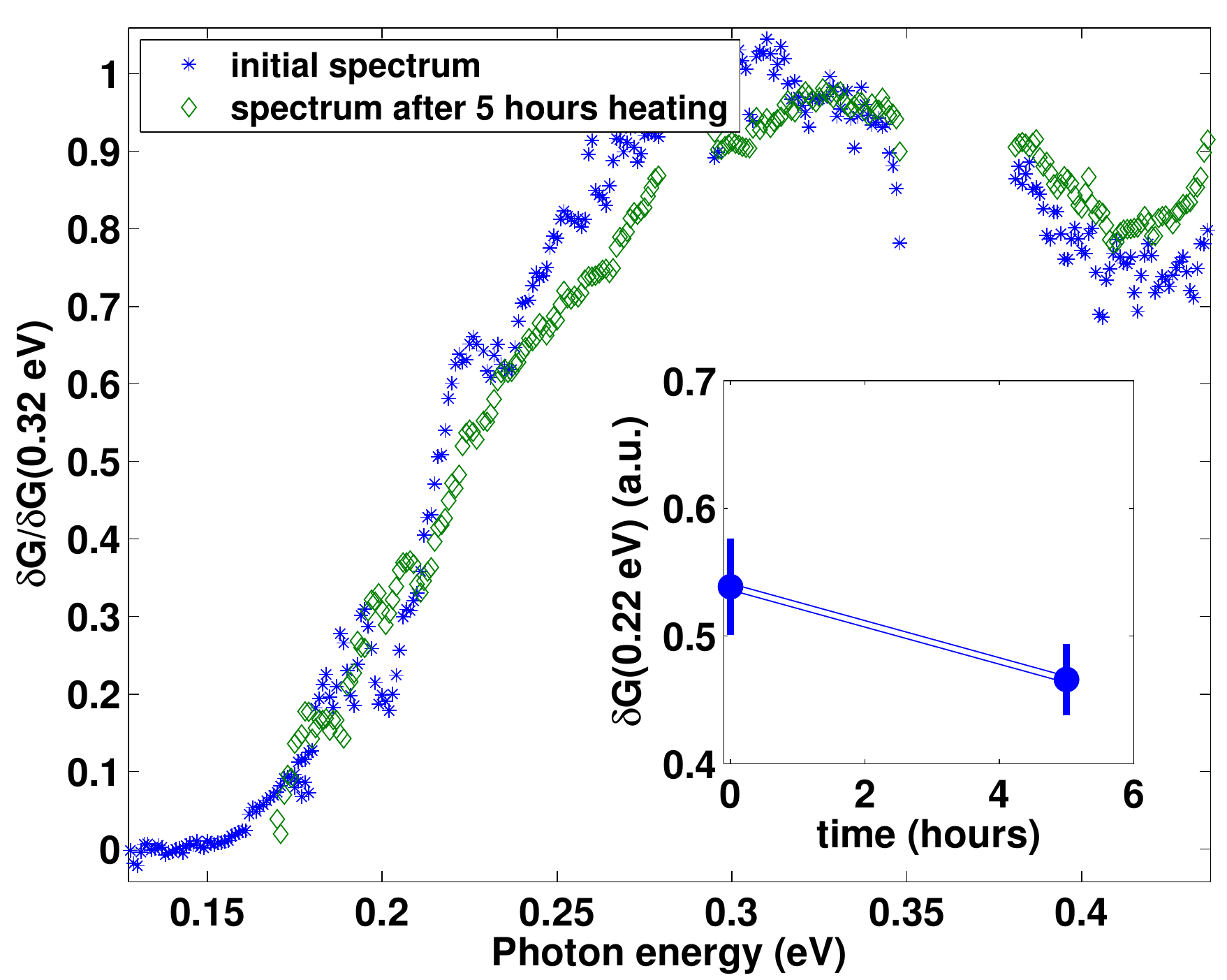}
\caption{Evolution of the photoconduction spectrum of the sample \#G with gold contacts after the same processing procedure. y-axes shows photoconductivity normalized by power of incident light, beside this all spectra are normalized to unity at maximum for convenience. $T=40$~K. Inset: the relative value of photoconductivity at 0.22 eV versus time of heating procedure}
\label{Fig.3}
\end{figure}

The curve plotted with stars on Fig~\ref{Fig.1} corresponds to the photoconduction spectrum of the sample \#Y etched in HF after 23 hours of processing. One of the In contacts was replaced  by a new one after the etching, so the resulted inter-contact distance $l$ becomes 30 microns shorter ($l_{new} = 357$~$\mu$m). No changes in the spectra associated with etching or contact replacing were observed.  
 However we detected small etching-induced decrease of the curvature of $\ln G(1/T)$ at the temperature range 50 - 100 K. Related type $G(T)$ change with decreasing of the sample thickness of {\it o-}TaS$_3$ whiskers was reported before \cite{zzmicro}.
The result of etching on $IV$ curves corresponds to the decrease of conductance, both linear and non-linear, obviously due to the cross-section area reduction. The absence of spectral changes is consistent with the statement that In is not confined to the surface \cite{Gill}.

Very similar growth of the low-energy contribution at $\hbar\omega < 0.25$~eV in the photoconduction spectra was found for another sample with In contacts after being kept for four years at room temperature in the air (Fig.~\ref{Fig.2}). The sample had almost the same length $l = 390$~$\mu$m as the sample \#Y. No such changes in the spectra were observed for much longer samples $l > 700$~$\mu$m. We interpret this as the evidence of slow In diffusion at room temperature that affects the spectra of the shortest samples. 

We present the photoconduction measurements results for the control sample \#G with the gold contacts before and after 5-hours heating on Fig.~\ref{Fig.3}. No significant change in the shape of the spectra  is seen at low photon energies $\hbar\omega < 0.25$~eV. However the very shape of the spectrum is different from the beginning. Namely the drop of the photoconduction is shifted to lower energies and smeared out compared to the sample \#Y with indium contacts. The  sample \#G has significantly lower threshold field for the CDW sliding $E_{th} = 0.7$~V/cm at 100 K than sample \#Y ($E_{th} =4.7$~V/cm) and thus \cite{monceau} lower impurity content (respective impurity concentration  $n_i \lesssim 10^{-4}$). From this fact and the polarization study of the same shaped spectra of pure samples \cite{tobepub} we suggest that such a spectral shape change may be a sign of the excitonic states seen in very pure samples of semiconductors and some 1D materials see for example \cite{pnas,averk}. Nevertheless, no new photo-conductive states appear in the spectra after the heating thus proving that evolution of photoconduction spectra presented on the Fig~\ref{Fig.1} is due to indium introduction.

\subsection{Linear conduction}
\begin{figure}
\includegraphics[width=7cm]{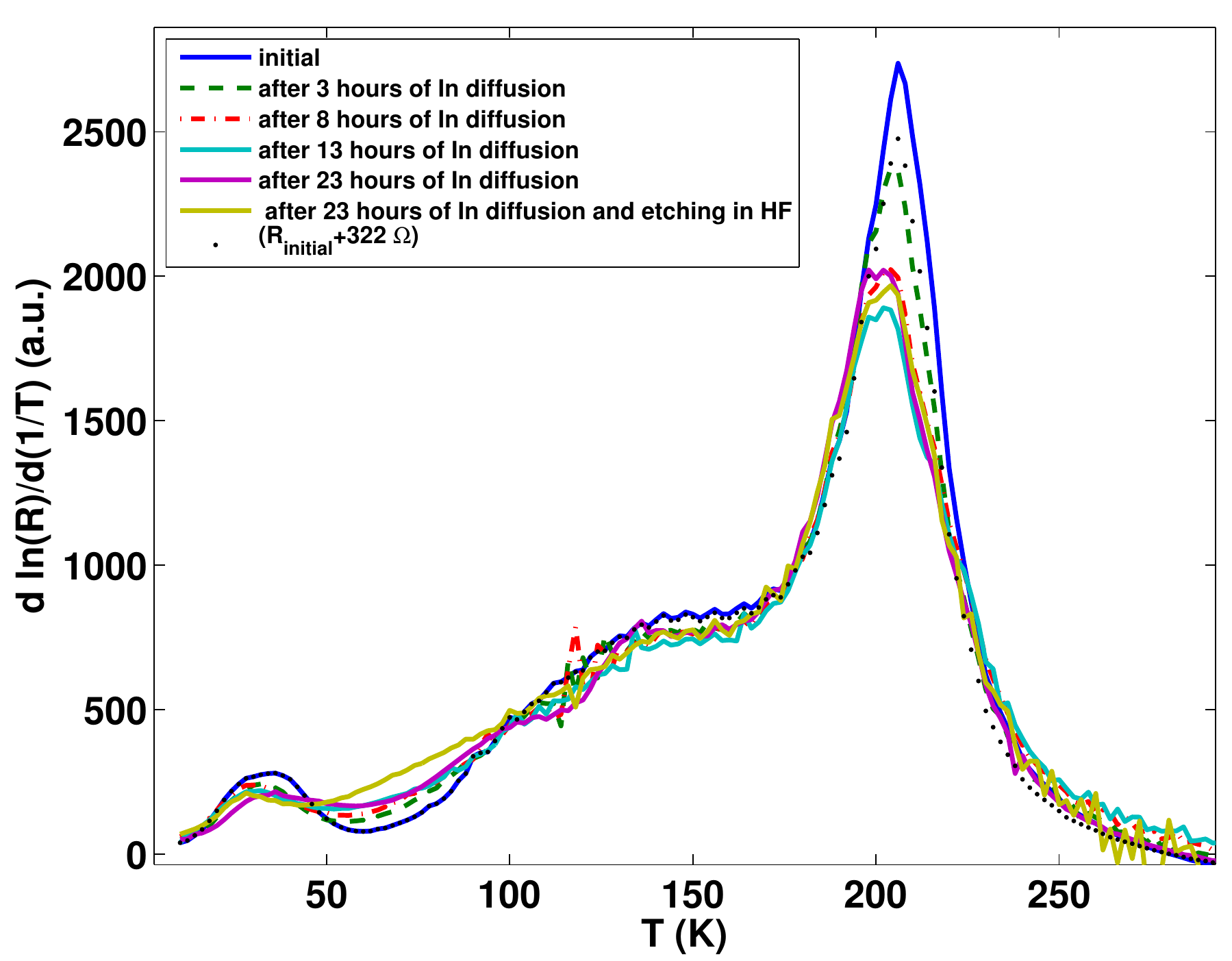}
\caption{The change of the $T_P$ with elongation of the In processng, revealed as the shift in maxima of logarithmic derivative of resistance dependence on temperature. Point curve correspond to the fitting  $G(T) = (G_0(T)^{-1}+\delta R_{300})^{-1}$ }
\label{Fig.rt}
\end{figure}

We observed that the Peierls transition temperature, $T_P$, for sample \#Y decreases with elongation of the heating processing and saturates after 8 hours of the processing (Fig.~\ref{Fig.rt}). The total change in $T_P$ was found to be $\delta T_P = -5$~K, the initial $T_P$ being equal to 206~K, the room-temperature resistance increased by $\delta R_{300} = 322~\Omega$ after 23 hours of the total processing time. Fig.~\ref{Fig.rt} shows also the data corresponding to $G(T) = (G_0(T)^{-1}+\delta R_{300})^{-1}$. Obviously,  $T_P$ changes can not be explained by the contact resistance change. Contrary, similar fit works well for the control sample with the gold contacts, perfectly matching both $G(T)$ and the logarithmic derivative (for this sample $\delta R_{300} = 252{\rm \ }\Omega$). No change in $T_P$ was observed for this sample. 

\subsection{Nonlinear conduction}
\begin{figure}[h]
\includegraphics[width=7cm]{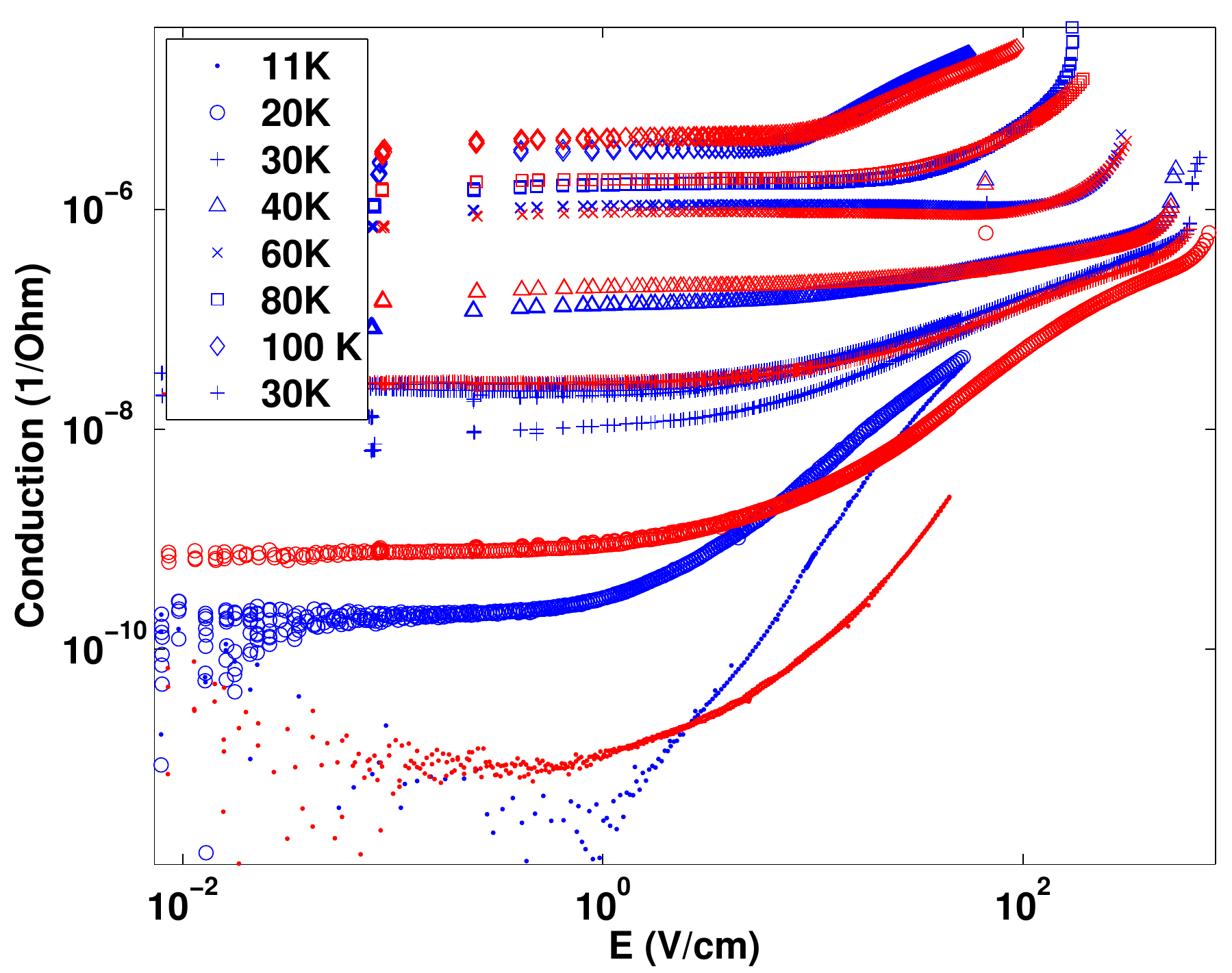}
\caption{The set of voltage-current characteristics for \#Y sample obtained after preparation (dark-blue curves) and after 8-hours of In processng (red curves). The same temperatures are plotted with same symbols}
\label{Fig.5}
\end{figure}

\begin{figure}
\includegraphics[width=7cm]{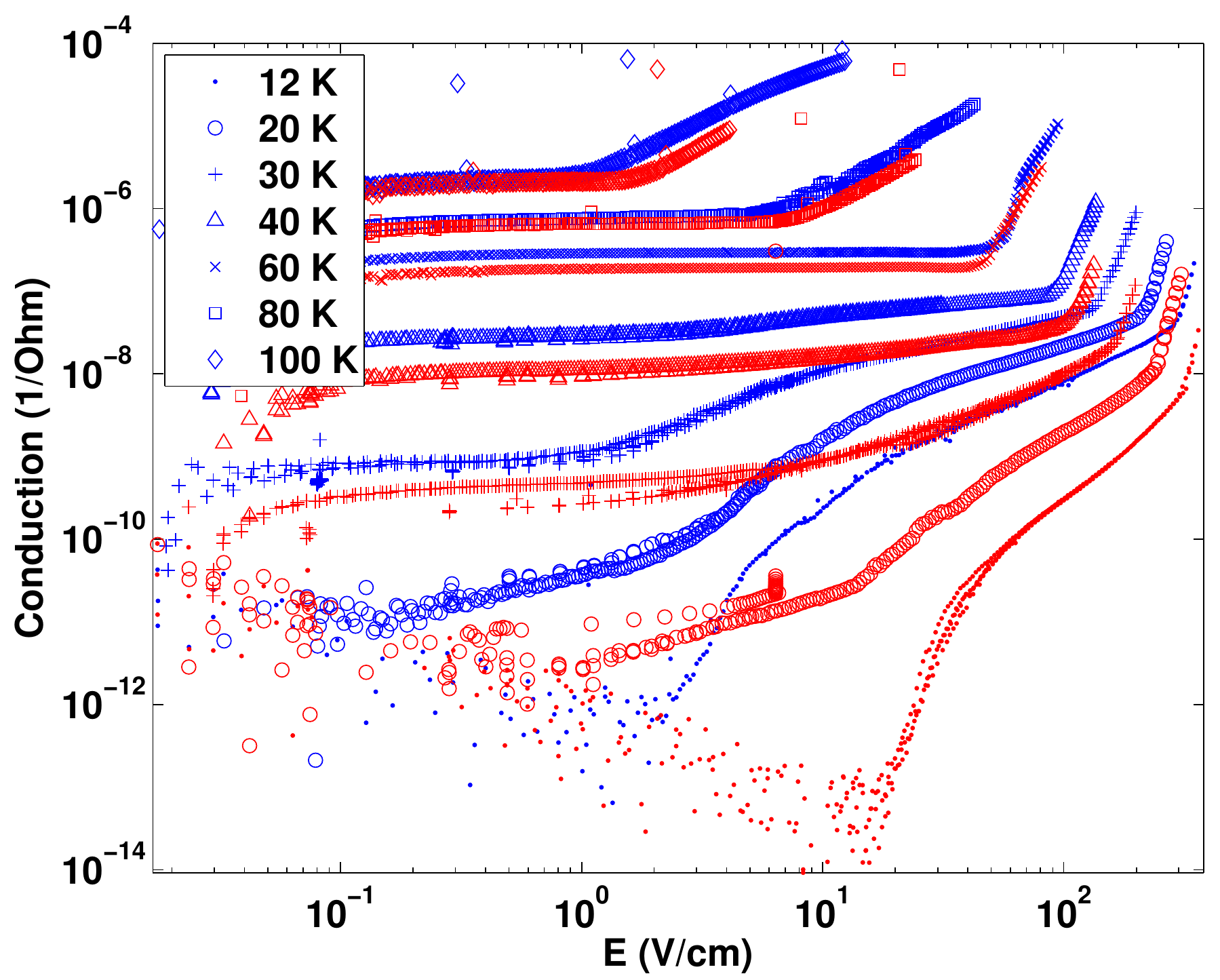}
\caption{The set of voltage-current characteristics for \#G sample obtained after preparation (dark-blue curves) and after 5-hours of temperature processing (red curves). The same temperatures are plotted with same symbols}
\label{Fig.6}
\end{figure}

Fig.~\ref{Fig.5} shows the initial set of $IV$ curves together with that obtained after 8 hours of processing for the sample \#Y. Respective sets for the control sample \#G is shown on Fig.~\ref{Fig.6}. The character of processing-induced $IV$ curves changes is apparently different. A shift in $E_{th}$ and suppression of both linear and non-linear conductance are observed for the sample \#G.
On the other hand, 8-hours processing of the sample \#Y results in the increase of the linear conductance and decrease of the non-linear one. The signature of such a behavior is the intersection of $IV$ curves obtained at the same temperatures before and after the processing. This is clearly seen at $IV$ curves at low temperatures, up to 40 K, but the same tendency exists up to 100 K. The decrease of the non-linear conductance with increasing In concentration is exactly the type of behavior one would expect due to the growth of the CDW pinning.  As for increase of the linear conductance at $T < 40$~K we believe it arises from the contribution of In impurities states providing additional conductance channel.

Increase of the linear conductance and decrease of the non-linear one  saturate when the total processing time  $t_{tot}$ exceeds 13 hours. Further processing leads to the changes which are similar to ones observed in the sample \#G, i.e. to the suppression of both linear and nonlinear conductivity. Meanwhile, the change in $T_P$ saturates when $t_{tot}$ is about 8 hours. This saturation defines the moment of the beginning of degradation of the sample and smearing out of the Peierls gap edge seen in Fig.~\ref{Fig.1}.

\subsection{Indium concentration distribution}
The concentration of added indium was measured using an electron probe micro analyzer as a function of the distance to the contacts, $L$, after 23-hours processing and sample etching in HF being completed. The results are presented in the table \ref{table}. The concentration of In within the $L = 5~\mu$m distance from contact is very high. It drops rapidly and becomes smaller than the sensitivity of the method ($\approx 0.3$~at.\% $ =  3 \times 10^{-3} $) at $L > 30~\mu$m. Such a distribution confirms diffusion of In from the contacts into the sample.

\begin{table}
\caption{\label{table} Distribution of In concentration after 23 hours of annealing at $T=120^o$C }
\begin{center}
\begin{tabular}{||c|c|c||}
\hline $L$, $\mu$m & Concentr., wt\% &  Concentr., at\% \\
\hline
\hline 5 & 19.58 $\pm$ 1.18 & 6.63 \\% \hline
\hline 7.5 & 7.26 $\pm$ 1.15 & 2.02 \\% \hline
\hline 10 & 2.19 $\pm$ 0.7 & 0.55 \\% \hline
\hline 31 & 1 or less & 0.3 or less \\% 
\hline
\end{tabular}
\end{center}
\end{table}

% The sensitivity was around 1 wt\%  that corresponds approximately to 0.3 at\%. 

A rough  estimate of In concentration can be done from the observed shift of $E_{th}$ at 100 K (see $IV$ curves and Fig.\ref{Fig.5}). The concentration of impurities in nominally pure {\it o-}TaS$_3$ samples is $n_0 = 10^{-4}$ per atom \cite{monceau}. In the weak pinning model $E_{th}(n_i) \propto n_i^\alpha$, where $\alpha$ varies between 2 and 2/3 for 3D and 1D pinning respectively. $E_{th}$ for pure samples $E_{th}(n_0) \sim 0.5$~V/cm. Initial threshold field for our sample \#Y was $E_{th} = 4.7$~V/cm. The respective initial impurity concentration can be estimated as $3\times 10^{-4}$ (3D pinning) - $3\times 10^{-3}$ (1D pinning). 

Taking $\delta E_{th}/E_{th(n_i)} = 0.2$ we observe after 23 hours of heating one gets $\delta n_i/n_i = 0.45$ for 3D case and thus the absolute concentration of introduced impurities is $\delta n_i \sim 1.2 \times 10^{-4}$, one order below the resolution of the electron probe microanalysis. For 1D pinning we obtain correspondingly $\delta n_i \sim 2.7 \times 10^{-4}$ that is still below resolution. The sample thickness $d \sim 1~\mu$m is more corresponding to the 3D pinning case.

Comparing this results to the results for NbSe$_3$ \cite{Gill}, we use $E_{th}(n_0 = 10^{-4}) \sim 0.1$~V/cm for pure samples of NbSe$_3$ at 115 K; most samples of NbSe$_3$ studied in \cite{Gill} $E_{th(n_0)}$ was of the same magnitude.
The average $E_{th}$ change observed after 10-60 minutes of In processing in \cite{Gill} was $\delta E_{th}/E_{th(n_0)} = 0.3$. Thus the same estimates as above give $\delta n_i = 0.9 \times 10^{-4} \sim 10^{-3}$. So the impurity concentration change for NbSe$_3$ samples is approximately one order bigger $\delta n_i$ than for our TaS$_3$ sample. This corresponds to the studies of In diffusion in tantalum dichalcogenides \cite{TaS2}, where it was found that for the sulfide TaS$_2$ diffusion coefficient was of the order lower than for the selenide TaSe$_2$. However diffusion coefficients for dichalcogenides seem to be much bigger than for trichalcogenides. 

Note that the measured concentration of In varies strongly with the distance to the contact, whereas the spectra does not change significantly when one of the contacts is replaced by another one in a new position which is $30~\mu$m closer. This indicates that the main contribution to the photoconduction comes from the regions of the sample with low In concentration rather than from near-contacts regions where the order parameter of the CDW is likely to be suppressed.

\section{Conclusion}
Our results clearly demonstrate that photoconduction of {\it o-}TaS$_3$ in the spectral region $ 0.15 < \hbar\omega < 0.25$ is not associated with the direct transitions between conduction and valence bands of the CDW gap. Instead, they depend on impurities content and probably structure defects created during impurities introduction. Such type of processes could be phonon-assisted transitions from the levels created by impurities near the bottom of the conduction band or the top of the valence band. Another type of the processes could be transitions associated with the levels created around structure defects due to local change of the gap value in their vicinity like those predicted in \cite{tutto}.
Thus the optical gap value of {\it o-}TaS$_3$ is actually $2\Delta = 0.25$~eV. The present results clarify the origin of the discrepancies in photoconduction spectra of the {\it o-}TaS$_3$ produced by R.E.~Thorne and F.~Levy \cite{TaS3} and that now can be associated with different impurities content. 

The observation that introduction of the impurities leads to the continuous tail in photoconduction spectra rather than to single peaks, consistently with previous results on Nb-doped samples \cite{ecrys2011}, is of particular interest. At the same time the peaks can be observed in pure samples of {\it o-}TaS$_3$ \cite{TaS3}. Thus the question on the origin of the electronic states corresponding to such single peaks remains open. Both the reported data on electric field dependence of the amplitude of such peaks and the new data \cite{tobepub} on polarization spectral studies of the TaS$_3$ bring strong evidences of the crucial role of the electron-electron and electron-phonon interaction at the origin of this states.

In summary, we observed In-dependent change in the photoconduction spectra and electrical properties of the {\it o-TaS$_3$}. Our results show that the optical gap value of {\it o-}TaS$_3$ is actually $2\Delta = 0.25$~eV.

\textbf{Acknowledgements.}
We would like to thank S.G. Zybtsev for gold contacts deposition and useful discussions, S.A. Brazovskii, T.V. and V.G. Krishtop for useful discussions. This work was supported by RFBR (projects \#12-02-31281 and \#14-02-01236) and Department of Physical Sciences of RAS.

% \appendix

% \section{}
% \label{}

\end{document}